\documentclass{article}
\usepackage[T1]{fontenc} % add special characters (e.g., umlaute)
\usepackage[utf8]{inputenc} % set utf-8 as default input encoding
\usepackage{ismir,amsmath,cite,url}
\usepackage{graphicx}
\usepackage{color}

\usepackage{lineno}

\newcommand{\yy}[1]{#1}
\usepackage{todonotes}

\title{Analysis and detection of singing techniques in repertoires of J-POP solo singers}

\multauthor
{{Yuya Yamamoto$^1$} \hspace{1cm} {Juhan Nam$^2$} \hspace{1cm} {Hiroko Terasawa$^1$}} { 
 $^1$ Doctoral Program in Informatics, University of Tsukuba, Japan\\
$^2$ Graduate School of Culture Technology, KAIST, South Korea\\
{\tt\small s2130507@s.tsukuba.ac.jp, juhan.nam@kaist.ac.kr, terasawa@slis.tsukuba.ac.jp}
}
\def\authorname{Yuya Yamamoto, Juhan Nam, and Hiroko Terasawa}
\begin{document}

\maketitle

\begin{abstract}
% Singing techniques in vocal performances are important and have close relationships to singers’ singing styles. Each professional singer of popular music has a unique singing style, that is, the use of singing techniques. 
 In this paper, we focus on singing techniques within the scope of music information retrieval research.
We investigate how singers use singing techniques using real-world recordings of famous solo singers in Japanese popular music songs (J-POP). 
First, we built a new dataset of singing techniques. 
The dataset consists of 168 commercial J-POP songs, and each song is annotated using various singing techniques with timestamps and vocal pitch contours.
We also present descriptive statistics of singing techniques on the dataset to clarify what and how often singing techniques appear.
We further explored the difficulty of the automatic detection of singing techniques using previously proposed machine learning techniques. 
In the detection, we also investigate the effectiveness of auxiliary information (i.e., pitch and distribution of label duration), not only providing the baseline. 
\yy{The best result achieves 40.4\% at macro-average F-measure on nine-way multi-class detection.}
We provide the annotation of the dataset and its detail on the appendix website{\protect\footnotemark[0]}.
\footnotetext[0]{\url{https://yamathcy.github.io/ISMIR2022J-POP/}}
\end{abstract}

\section{Introduction}\label{sec:introduction}
A singing voice is one of the most essential elements of music. It provides impactful emotional expressions through melody and lyrics. 
In particular, in popular music, the role of the singing voice is more important, as the vocal quality and unique style of singers are critical in attracting people.
Vocals mainly consist of singer individuality (i.e., vocal fold vibration and vocal tract resonance) and singing expressions (i.e., fine control of pitch, timbre, and loudness). Singing techniques, as the name suggests, are extended techniques in human singing. % (i.e., the same as the playing techniques in instrumental performances)
It characterizes singing voices as a component of expressions in vocal performances.
In particular, many professional singers in popular music use singing techniques to make a performance more expressive, each in a different manner.

The singing voice is an important subject in the field of music information retrieval (MIR), and there are many prior works that extract quantitative or objective information as in the task of singing transcription\cite{Ryynanen2006}, lyric identification\cite{humphrey2018introduction}, and singer identification\cite{humphrey2018introduction}. 
Computational analysis of singing techniques based on MIR can contribute to the clarification of singing styles and can be applied to music discovery, vocal training, consumer-generated media and so on.
Singing techniques have been of keen interest, especially in J-POP, both among singers and music creators; the singing style of J-POP singers has a wide variety, and their singing techniques are diverse. 
% In addition, Japanese commercial karaoke systems with scoring systems (e.g., DAM\footnote{\url{https://www.clubdam.com/}} from Daiichi-Kosho CO., LTD, Joysound\footnote{\url{https://www.joysound.com/web/}} from XSing INC., etc.) and singing techniques are one of the evaluation criteria. 
In addition, singing techniques are one of the evaluation criteria in Japanese commercial karaoke systems with scoring systems.
Meanwhile, as for music creators, \textit{VOCALO} songs, whose singer's voice are created by a singing voice synthesis software such as VOCALOID \cite{Kenmochi2007VOCALOID}, have been established as a music genre in Japan.
Many VOCALOID creators manually manipulate the voice of VOCALOID to make it more expressive\yy{, sometimes while referring to how the actual singer produce the singing voice, like how singing techniques are applied}.
Therefore, J-POP is an attractive research target for singing technique analysis and computational singing technique analysis may bring benefits to real-world applications.

In this regard, we present a case study of singing technique analysis in J-POP. 
We first introduce a new dataset, including the annotation of singing techniques for J-POP recordings. 
We analyze the dataset using descriptive statistics to illustrate the technique distributions and singer styles. Finally, we conduct singing technique detection based on machine learning methods\yy{, which can be beneficial to analyze the one side of the singing voice automatically}. 
We demonstrate the detection of various singing techniques (i.e., identifying the singing technique type with its starting time and duration) and report baseline results using existing methods. Finally, we investigate the effect of the identification model with auxiliary information derived from the statistics of the dataset and pitch information.

\section{Related works}

\subsection{Singing Technique Analysis}

Analyzing the acoustic patterns of singing expressions has been a long-standing research topic. Research on vibrato dates to the 1930s \cite{seashore1938musical}. Later, computer-based methods allowed a more quantitative analysis of vibrato parameters \cite{prame1994measurements, sundberg1999perception}. In addition to vibrato, other singing techniques, such as growling \cite{sakakibara2004growl}, portamento \cite{yang2016ava}, voice registers \cite{hirayama2012discriminant, lee2021differences}, and extreme vocal effects\cite{nieto} have also been investigated. 
An analysis of singing techniques was also conducted for J-POP. 
Migita et al. investigated the vibrato parameters of the imitated singing voices of J-POP vocalists \cite{migita2010study}.
\yy{Yamamoto et al. conducted an exploratory analysis of singing techniques that imitate the singing style of J-POP singers \cite{yamamoto2021analysis}. 
They found 13 types of singing techniques from 48 tracks and analyzed them with annotations. 
% relationships between singers and occurrence locations. 
Because singing techniques in J-POP cover wide repertoires\cite{akira}, such exploratory analysis is also important for the purpose of singing performance analysis. 
As our interests are handling and detecting multiple singing techniques rather than specifying a single technique,
we annotated the label as in the manner in \cite{yamamoto2021analysis} but on the original real-world tracks of J-POP.  
Therefore, our new dataset consists of labels for multiple singing techniques.}

\subsection{Singing Technique Datasets}

There are a handful of datasets that focus on the analysis of singing techniques. The Phonation Mode dataset consisted of four vocal modes (neutral, pressed, breathy, and flow) of sustained sung vowels from four singers \cite{proutskova2013breathy}. Although the dataset includes a wide range of pitches, they are discrete, and thus lack a melodic context. VocalSet \cite{wilkins2018vocalset} handles the issue by having voices sung in
contexts of scales, arpeggios, long tones, and excerpts. In addition, it covers a broader range of singing techniques, such as vibrato, trill, vocal fry, and inhaled singing. Because audio samples in the two datasets were newly recorded, they were collected under controlled conditions.  
\yy{Therefore, their characteristics of singing techniques might be different from those appeared in songs.}

However, several datasets have been annotated for real-world vocal music.
The KVT dataset was originally built for vocal-related music-tagging tasks in K-POP music \cite{kim2020semantic}. It contains 70 vocal tags, of which six are related to the singing technique (whisper/quiet, vibrato, shouty, falsetto, speech-like, and non-breathy). The annotation was conducted using crowdsourcing. The MVD dataset was built to analyze screams in heavy metal music, and has four different types of screams (high fry, mid fry, low fry, and layered) \cite{kalbag2022scream}. The regions and types of screams in the audio files were manually annotated. Similar to these studies, we are interested in versatile singing techniques in real-world music and thus we chose commercial music to build the dataset.    
%\yy{Considering all of these works, our interest lies more on versatile singing techniques that are appeared in real-world song pieces.}

\subsection{Singing/Playing Technique Identification on MIR research}

Studies have been conducted on the automatic identification of singing techniques beyond computational analysis. There are two scenarios for identifying a singing techniques. One is classifying sung vocal audio into singing techniques. Several studies have conducted singing technique classification on VocalSet\cite{yamamoto2021investigating} and phonation modes \cite{rouas2016automatic, stoller2016analysis}. These works identify singing techniques but do not provide time-related information, such as start time and duration.  
The other method is to detect the singing technique in time. 
Miryala et al. \cite{miryala2013automatically} identified the singing expressions of raga, classical music in India. They created 35 recordings in eight ragas sung by six singers and showed a classification accuracy of 84.7\% using a rule-based classifier. 
Yang et al. \cite{yang2016ava} proposed AVA, an interface for analyzing vibrato and portamento based on the filter diagonalization method (FDM) and hidden Markov model (HMM), respectively. They also provided a case study of the analysis of vibrato and portamento in the Beijing opera.
Ikemiya et al. \cite{ikemiyatransferring} provided rule-based parameterization of four singing expressions (vibrato, kobushi, gliss-up, gliss-down), extracted them from recorded vocal tracks to transfer to other vocal tracks, and demonstrated them on two sung excerpts.
Kalbag et al. \cite{kalbag2022scream} provided scream detection for heavy metal music. They also built the MVD dataset and demonstrated classification, including non-scream singing and non-vocal music.
Since singing techniques appear locally on sung vocals, we adapted these temporal detection strategies for identification.

\begin{figure*}[t!]
    \centering
    \includegraphics[width=\textwidth]{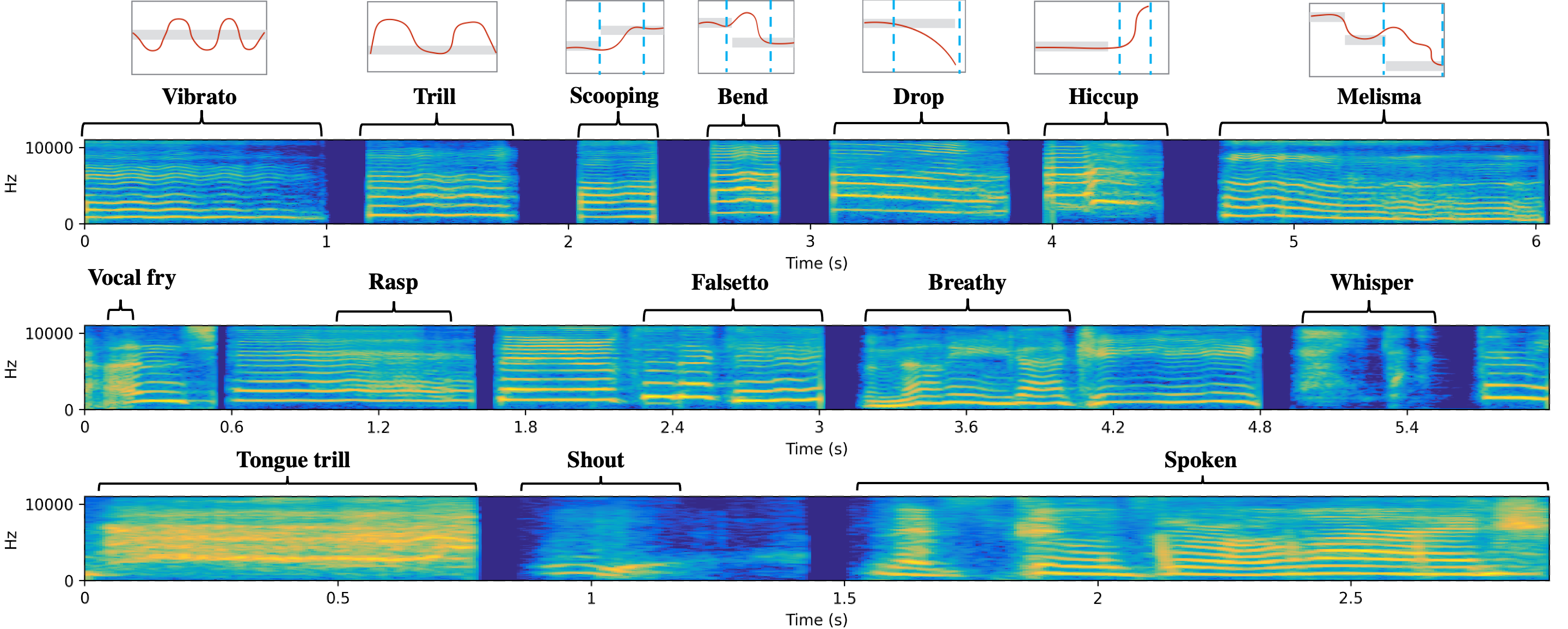}
    \caption{Singing techniques included in the dataset. (upper) Pitch techniques with a sketch of pitch contour. Gray is the target vocal note, the red line is the pitch contour and blue dotted lines are boundary of each technique. (middle) Timbral techniques. We also show the non-technique extracted from the same track. (lower) Miscellaneous techniques.}
    \label{fig:overview}
\end{figure*}

\section{Dataset}
In this section, we describe the organization of our new dataset used for the analysis of this study. 
Due to the limit of the space, more details are described in the appendix site{\protect\footnotemark[0]}.
%To enable this study, as well as to facilitate future research on the singing technique analysis of popular music, we decided to build a new dataset.
\subsection{Song Selection}
To cover a wide range of singing techniques, the dataset should include various types of vocalists, considering gender, genre, tempo, and mood.
%We prioritized them including various songs in our dataset over multiple fully annotated verses in a song to collect diverse singing styles.
We first listed 42 solo singers (21 males and 21 females), and four famous hit songs were selected from each singer to be as different as possible. Each song was performed as solo performance in Japanese.
We collected audio tracks from commercial CD recordings of the J-POP songs.
% We prioritized including various songs in our dataset to fully annotate multiple verses in a song to collect diverse singing styles. 
\yy{We trimmed the collected audio tracks and annotated only the first consecutive section (i.e., verse-A, verse-B and Chorus).
Since we prioritized including various songs in our dataset rather than fully annotating multiple verses in a single song, we could collect a diverse set of singing styles.}

When using popular commercial songs for academic research, copyright is always an issue. The works by Kim \cite{kim2020semantic} on K-POP and Kalbag \cite{kalbag2022scream} on heavy metal music are good examples of developing datasets using commercial music, which distributes only labels about the performed music but not the audio signals or score information. We followed their solutions and made our dataset that consists of singing technique and pitch (frequency) contour labels. 

% \subsection{Annotation methodology}
\subsection{Data Pre-processing}
The dataset contains two types of annotations: vocal melody and singing techniques.  We illustrate the annotation result in Figure \ref{fig:sonic_visualiser}. All annotations were conducted on isolated vocal tracks after vocal separation using Demucs v3 \cite{defossez2021hybrid}, which is a state-of-the-art model for musical source separation. During the annotation process, we confirmed that there was no dropout of vocal regions by comparing the original mixed track. 
Instead, we observed that the separated vocals tends to retain the back-chorus or sometimes instrumental sounds that are similar to the singer's voice (e.g., electric guitar, synthesizer).
\yy{
\begin{figure}[!t]
    \centering
    \includegraphics[width=\columnwidth]{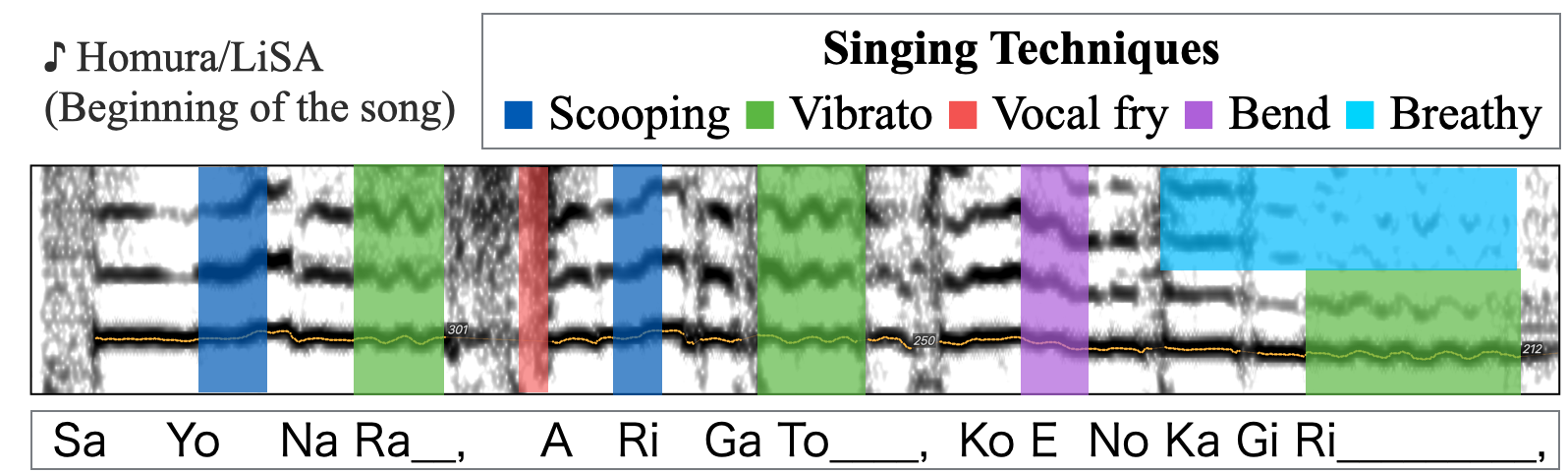}
    % \caption{Screen shot of vocal melody (orange curve) and singing technique (horizontal red bar) annotation on melodic range spectrogram, visualized by Sonic Visualiser.}
    \caption{An overview of annotation. (middle) an excerpt spectrogram with singing technique (colored area) and vocal melody (orange curve).}
    \label{fig:sonic_visualiser}
\end{figure}
}

\subsection{Singing Technique Annotation}

\begin{table}[h!]
\centering
\label{tab:description}
\scalebox{0.9}{
\begin{tabular}{|l|l|l|}
\hline
Technique      & description                                                                                       & type          \\ \hline \hline
vibrato        & a periodic oscillation of pitch.                                                                  & pitch         \\ \hline
scooping       & an upper continuous picth change                                                                  & pitch         \\ \hline
drop           & a lower continuous picth change                                                                   & pitch         \\ \hline
bend           & \begin{tabular}[c]{@{}l@{}}a short tremolo or U/inverted-U \\  shaped pitch change\end{tabular} & pitch         \\ \hline
hiccup         & \begin{tabular}[c]{@{}l@{}}a short hiccuping \\ on attack/release of note\end{tabular}            & pitch         \\ \hline
melisma &
  \begin{tabular}[c]{@{}l@{}}a musical arrangement \\ in which several notes are \\ applied to one syllable of a lyric.\end{tabular} &
  pitch \\ \hline
trill          & \begin{tabular}[c]{@{}l@{}}a continuous pitch change \\ between two notes\end{tabular}            & pitch         \\ \hline
falsetto       & sung by falsetto register.                                                                        & timbre        \\ \hline
breathy        & sung by breathy sound.                                                                            & timbre        \\ \hline
whisper        & sung like whispering.                                                                             & timbre        \\ \hline
rasp           & \begin{tabular}[c]{@{}l@{}}sung by a creaky voice \\ with subharmonics.\end{tabular}              & timbre        \\ \hline
vocal   fry    & \begin{tabular}[c]{@{}l@{}}sung by a creaky voice \\ and pulse register phonation.\end{tabular}   & timbre        \\ \hline
spoken &
  \begin{tabular}[c]{@{}l@{}}singing like rapping, \textit{sprechgesang}\footnotemark[1],\\ and some other styles like speaking.\end{tabular} &
  misc. \\ \hline
shout          & shouting.                                                                                         & misc. \\ \hline
tongue   trill & \begin{tabular}[c]{@{}l@{}}a rolling tongue, \\ occurred on {[}r{]} consonant.\end{tabular}       & misc. \\ \hline
\end{tabular}
}
\caption{The description of each singing technique appeared in the dataset.}
\end{table}

\footnotetext[1]{A vocal style between singing and speaking.}

% We annotated singing techniques exploratorily.
We thoroughly surveyed singing techniques based on instructional books \cite{akira, husler1976singing} and other conventional scientific research related to singing techniques \cite{wilkins2018vocalset, kim2020semantic, ikemiya, Panteli2017towards, sakakibara2004growl, nakazato, yamamoto2021analysis, jones2016rhythm} and defined the labels for the major techniques that appear in these references. We manually annotated songs using these labels. 
 
Although many other singing techniques still exist\footnotetext[2]{Although discarded in the analysis, we also annotate `unknown' labels if the region seems to represent a singing technique but is difficult to classify them into any of the techniques above and found 24 unknown techniques in total. In the techniques, there are some sounds akin to coughing and pig squeals, for example.}, we considered the following 15 singing techniques in this study. 
We show the description of each singing technique in Table 1. 
The pitch contour sketches of these techniques and spectrogram examples are illustrated in Figure \ref{fig:overview}. 
 
% \begin{itemize}
%     \item {\textbf{Category 1: Pitch techniques}}
%     \begin{itemize}
%     \item \textbf{vibrato}: a periodic oscillation of pitch.
%     \item \textbf{scooping}: a pitch evolution like upward glissando.
%     \item \textbf{drop}: a pitch evolution like downward glissando.
%     \item \textbf{bend}: a short tremolo or U/inverted U-shaped pitch evolution.
%     \item \textbf{hiccup}: a short hiccupping on attack/release of note.
%     \item \textbf{melisma}: a musical arrangement in which several notes are applied to one syllable of a lyric.
%     \item \textbf{trill}: a continuous pitch change between two notes.
%     \end{itemize}

%     \item {\textbf{Category 2: Timbre techniques}}
%     \begin{itemize}
%     \item \textbf{falsetto}: sung by falsetto register.
%     \item \textbf{breathy}: breathy sound.
%     \item \textbf{whisper}: sung like whispering.
%     \item \textbf{rasp}: sung by a creaky voice, with subharmonics.
%     \item \textbf{vocal fry}: sung by a creaky voice and pulse register phonation.
%     \end{itemize}

%     \item {\textbf{Category 3: Miscellaneous techniques}}
%     \begin{itemize}
%     \item \textbf{spoken}: singing like rapping, \textit{sprechgesang} (i.e., a vocal style between singing and speaking.), and some other styles like speaking.
%     \item \textbf{shout}: shouting.
%     \item \textbf{tongue trill}: a rolling tongue, accompanied with [r] consonant.
%     \end{itemize}
% \end{itemize}

We note that these label names are not unique in the real-world (e.g., scooping is also called `portamento,' `glissando,' `gliss-up,' `shakuri' (in Japanese)).  
We made frame-level annotations on the audio tracks collected based on the vocabulary. 
\yy{Singing techniques were carefully annotated by an experienced vocalist (i.e., the first author of the paper) with the helps of sound playback and visualizing the spectrograms and pitchgrams.
The annotation process is conducted on Sonic Visualizer \cite{cannam2010sonic}.}
% Figure \ref{fig:sonic_visualiser} shows a screenshot of the annotation on the editor of the Sonic Visualizer. 

\subsection{Melody Annotation}
\label{sec:melody}
Since pitch is an essential component of singing technique analysis, we further annotated melodic pitch using Tony \cite{mauch2015computer}, followed by manual correction such as removing the unvoiced parts and reverberation tails.

\section{Descriptive Statistics} 

% In this section, we show the descriptive statistics of the dataset to summarize the singing techniques in J-POP.
\subsection{Song Statistics}

\begin{figure} [!t]
    \centering
    \includegraphics[width=\columnwidth]{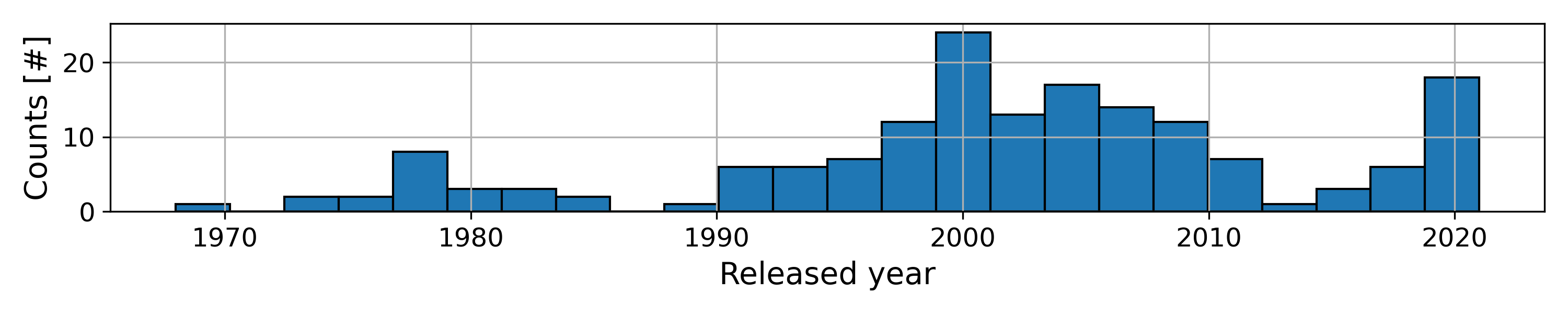}
    \caption{Distribution of song year of release.}
    \label{fig:release}
\end{figure}

\begin{figure} [!t]
    \centering
    \includegraphics[width=\columnwidth]{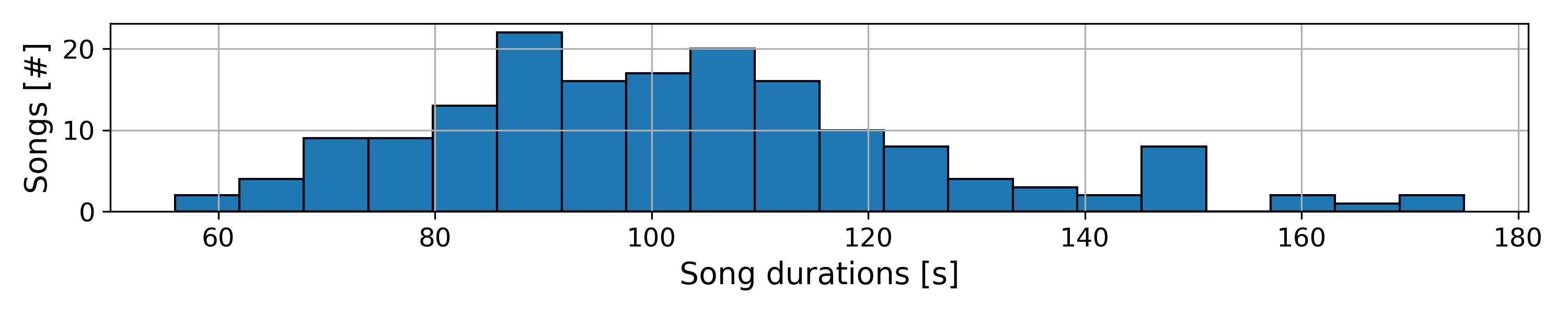}
    \caption{Distribution of song length in seconds.}
    \label{fig:length}
\end{figure}

The distribution of songs selected for the dataset based on the year of release is shown in Figure \ref{fig:release}. Songs can be collected from various eras, ranging from 1968 to 2021. The distribution of the songs in the dataset is shown in Figure \ref{fig:length}.
The overall length was 4h 47m 39s, and the average length of a song track was 1m 43s. The ratio of technique regions per song track was 22.8\%.

\subsection{Label Statistics}
\label{sec:analysis}
 %Next, 

We present the statistics for the annotated labels as a histogram at the upper side of Figure \ref{fig:stats}. The most frequent technique is `scooping’. It is followed by `vibrato’, `bend’, and `drop’. This indicates that such techniques are relatively common in J-POP. These techniques are also used in Japanese commercial karaoke systems for vocal assessment\cite{karaoke}. `scooping’, `bend’ and `drop’ are portamento, whose frequent use is sometimes considered undesirable in classical singing \cite{husler1976singing}. It can be said that this frequent use of portamento is a characteristic of J-POP.
 
\begin{figure}[!t]
     \centering
     \includegraphics[width=\columnwidth]{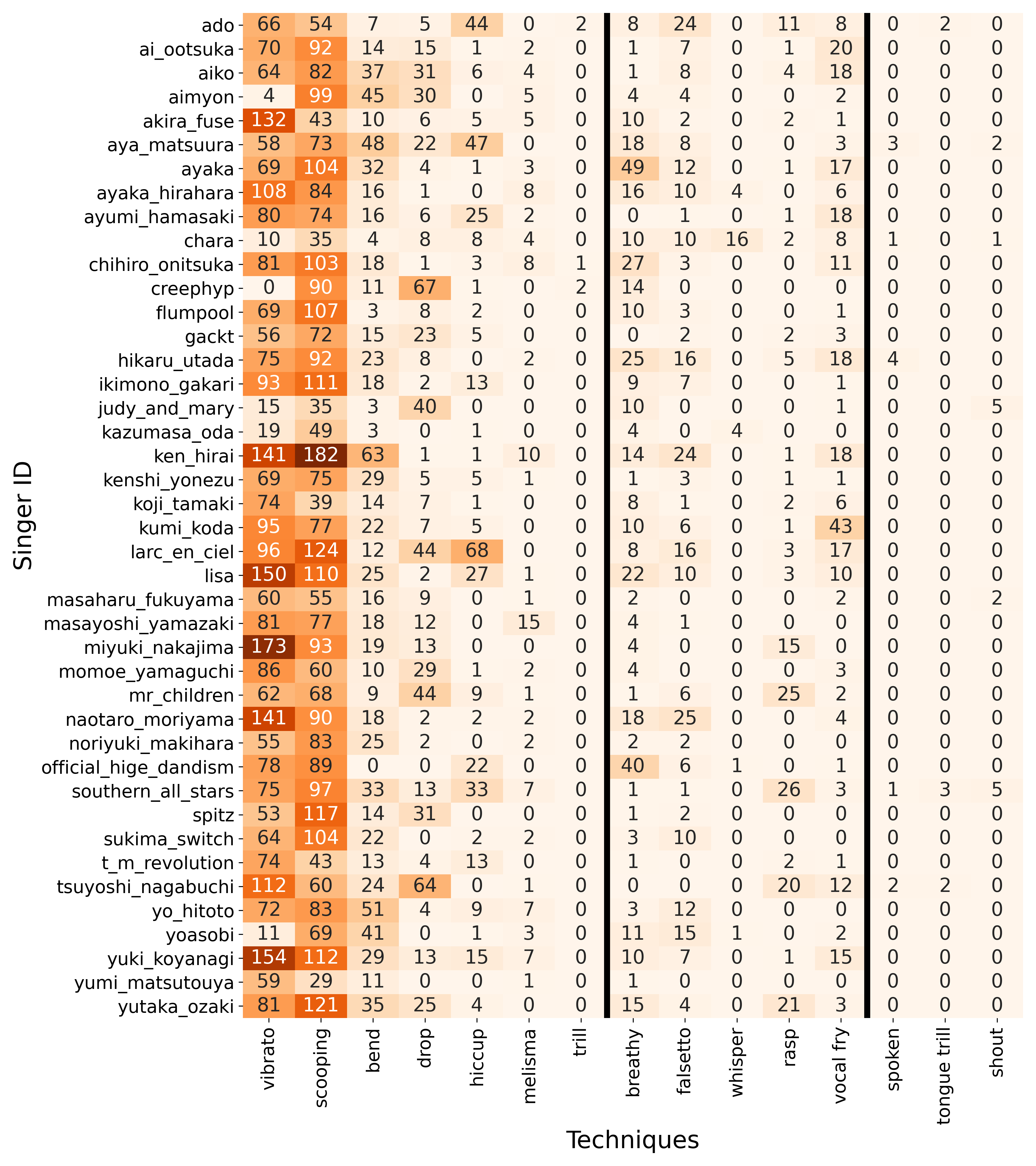}
     \caption{Occurrence distribution of singing techniques by singer. The vertical black line divides each category.}
     \label{fig:singer-wise}
 \end{figure}

We show the distribution of techniques by each singer in Figure \ref{fig:singer-wise} and find that the occurrence frequency of the techniques is different for each singer. We also confirmed that scooping appears more than 29 times for every singer, whereas several singers tend to not use vibrato frequently (e.g., `creephyp’, `aimyon’, and `yoasobi’). 
 
\begin{figure}[!hbt]
     \centering
     \includegraphics[width=\columnwidth]{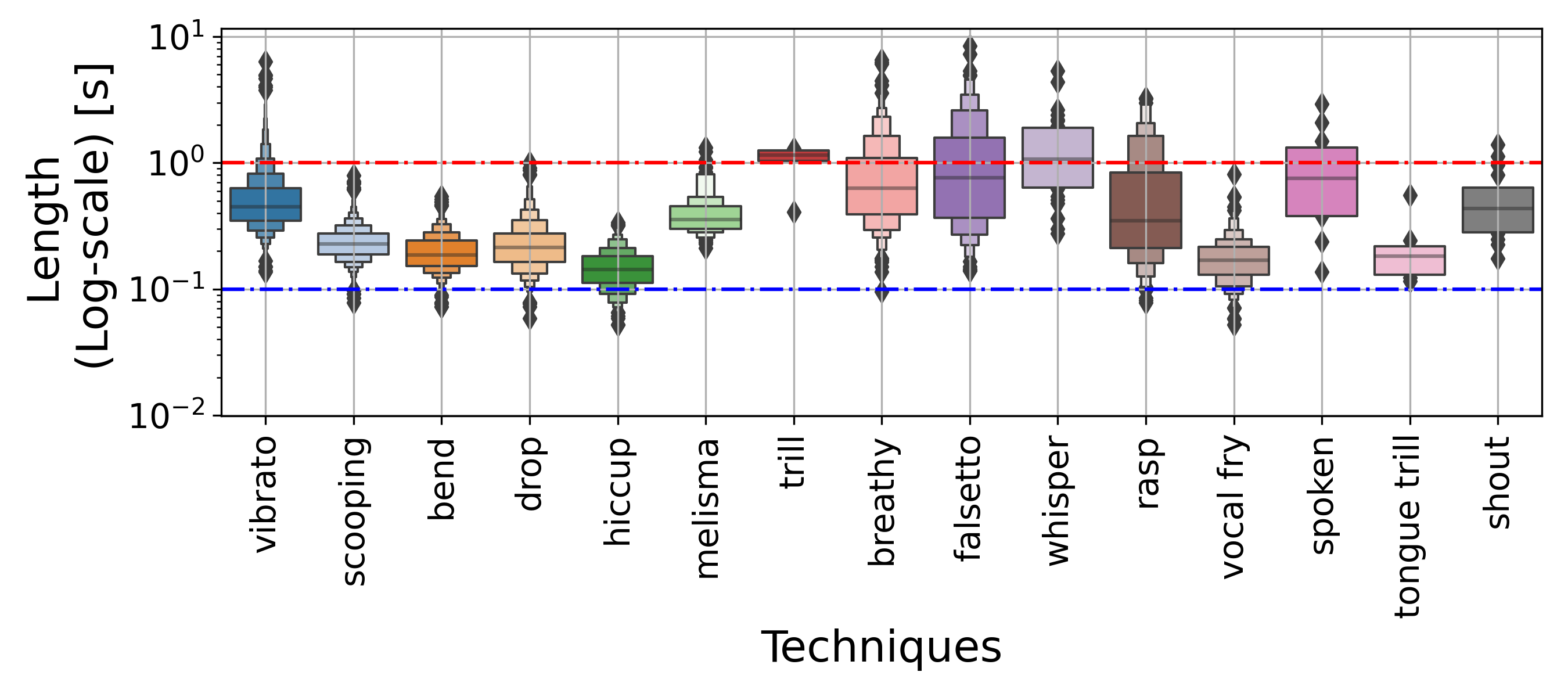}
     \caption{Distribution of each technique. Red and blue horizontal lines indicate 1 s and 0.1 s, respectively.}
     \label{fig:dist-duration}
 \end{figure}

The lower side of Figure \ref{fig:stats} and Figure \ref{fig:dist-duration} shows the total duration and distribution of each technique using a boxenplot\footnotetext[3]{https://seaborn.pydata.org/generated/seaborn.boxenplot.html}, respectively. Figure \ref{fig:dist-duration} shows that most of the techniques are \yy{relatively short} (that is, the range between 0.1s and 1s.), especially for `drops', `scooping', `bend', `hiccup', and `vocal fry'. The average length of the singing techniques was 0.4s. 
 
% \begin{figure}[!t]
%      \centering
%      \includegraphics[width=\columnwidth]{figs/count.png}
%      \caption{Technique counts.}
%      \label{fig:count}
% \end{figure}
 
%  \begin{figure}[!t]
%      \centering
%      \includegraphics[width=\columnwidth]{figs/duration.png}
%      \caption{Technique duration.}
%      \label{fig:duration}
%  \end{figure}
\yy{
 \begin{figure}[!t]
     \centering
     \includegraphics[width=\columnwidth]{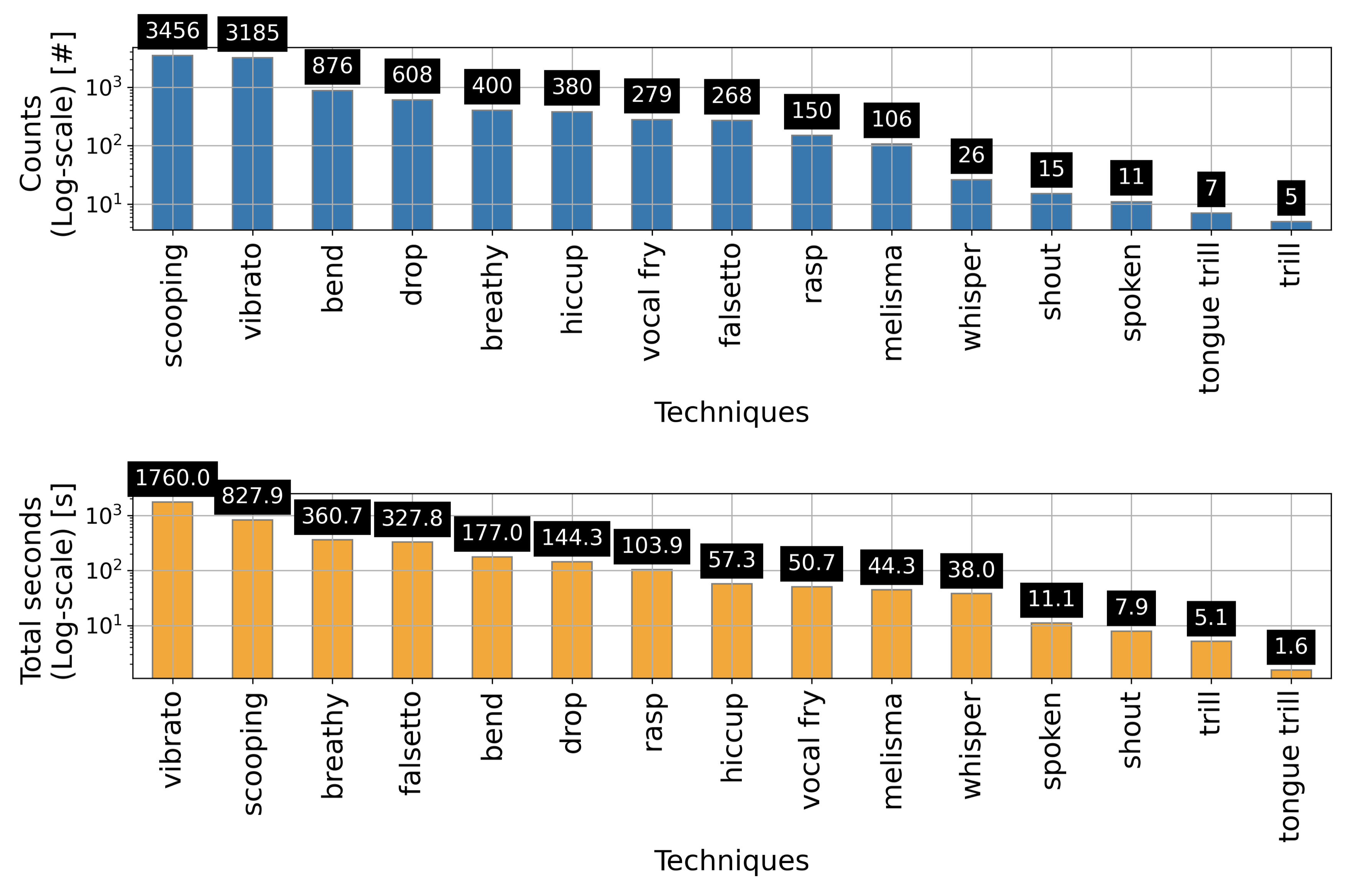}
     \caption{Statistics of the labels. upper: counts, lower: total duration.}
     \label{fig:stats}
 \end{figure}
 }

\section{Singing technique detection}
\begin{figure}[!t]
    \centering
    \includegraphics[width=\columnwidth]{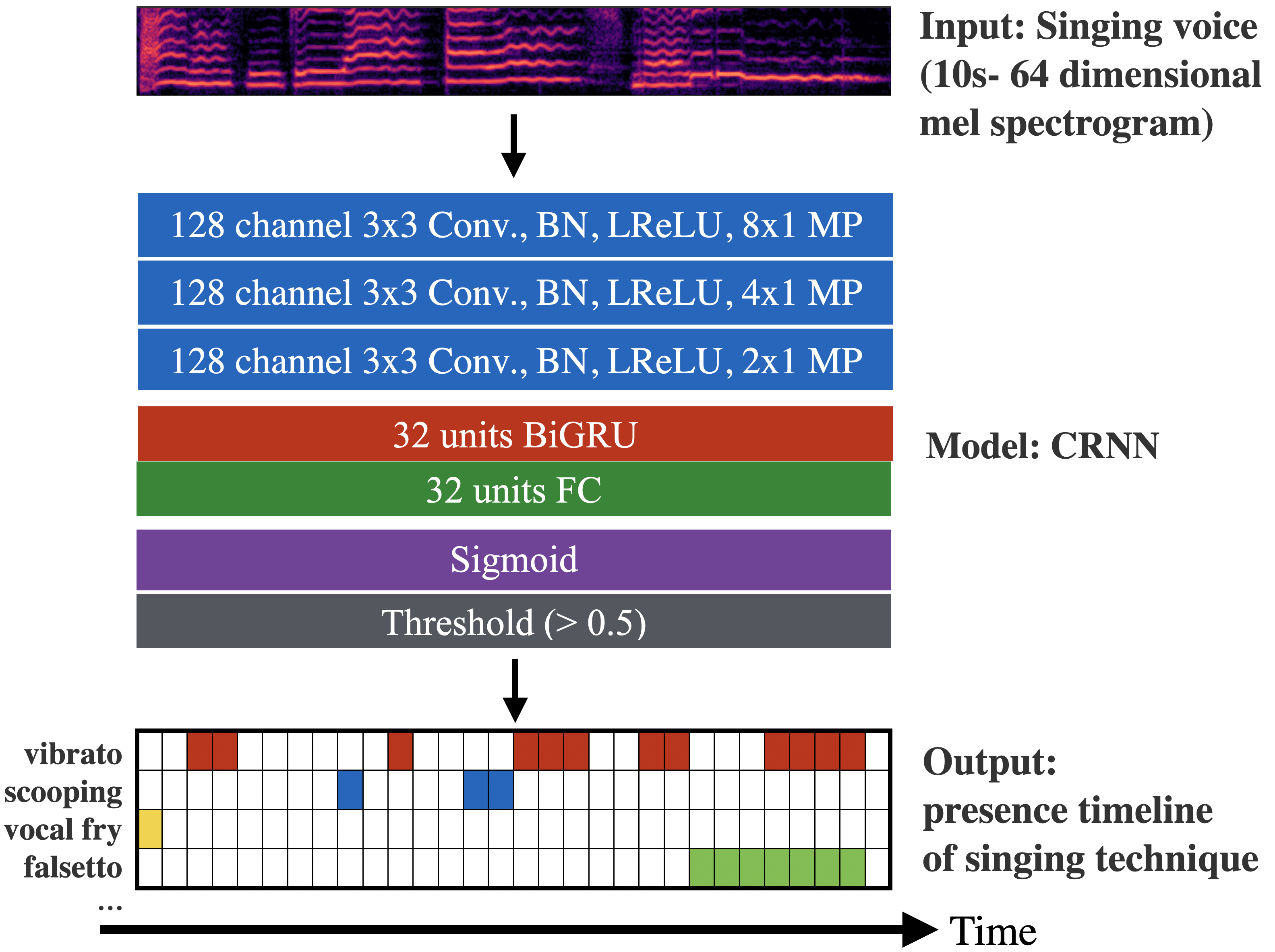}
    \caption{An overview of the singing technique detection. The configuration of the CRNN model is also shown.}
    \label{fig:detection}
\end{figure}

In this section, we describe the detection of the singing techniques.
We conducted experiments on a multi-class detection scenario, which handles classification and localization simultaneously. The problem setting is the same as that in sound event detection (SED) \cite{mesaros2021sound}.
Figure \ref{fig:detection} illustrates the proposed method. 
We investigated the performance of the CRNN model in a singing technique detection task. 
\yy{In addition to running the simple setting, we also investigated how the considerations of the characteristics of the dataset improve the performance (i.e., label sparseness and pitch information).}
Thus, through the experiment, we attempted to answer for the following three research questions: 
1) To what extent can we detect singing techniques using machine learning? 
2) Can modification of the loss function treat the problem of label sparseness? 
3) Can auxiliary pitch information help improve the detection performance? 

\subsection{Experimental Conditions}
We performed singer-wise seven-fold cross validation during the experiment.
% We split audio clips into a training set, validation set, and test set with a ratio of 5:1:1. 
\yy{We first split the singers into seven groups. 
Then, in each run, one group is left out for test set, another one group is for validation set , and the rest are used for training set.}
% Thus, no singers in the training and validation sets appeared in the test set. 
Owing to the label imbalances of techniques between singer, we used the 9 most common classes (i.e., `bend,' `breathy,' `drop,' `falsetto,' `hiccup,' `rasp,' `scooping,' `vibrato,' and `vocal fry'), which appear on every fold. We divided the singer fold to balance the amount of each technique as much as possible.

We segmented \yy{the vocal tracks, which are separated by Demucs v3 \cite{defossez2021hybrid} in advance,} into 10s audio clips and non-overlapping parts at a sample rate of 44.1 kHz. Therefore, they were converted into 64-dimensional log-mel spectrograms. 
For experiments that involve the computation of short-time Fourier transform (STFT), we used a Hann window with 2048 samples to compute the discrete Fourier transform (DFT). 
The hop size was set to 10 ms in every experiment.

We used a CRNN model, which is widely used as a baseline system for SED. 
\yy{We adapted the settings of the model of Imoto et al.\cite{imoto2021impact}, which treated a similar issue (i.e, label sparseness) of ours in SED. }
The model configuration is shown in Figure \ref{fig:detection}.
The model consists of three convolutional layers, one layer of bidirectional GRU cells, and one fully connected (FC) layers. 
The input was the aforementioned log-mel-spectrogram, and the output was a multi-hot vector representation of the singing technique for each time frame. 
All experiments were conducted using a batch size
of 16, and the model parameters were optimized using
an Adam optimizer with a learning rate of $10^{-4}$. The model stopped training if the value of the loss function on the validation set did not improve over 10 epochs.

The model was optimized using the binary cross-entropy (BCE) loss.
% as follows: 
% $$L_{bce}(p_t) =  -((1-p_t)\log(1-p_t)+p_t\log(p_t))$$ 
% where $p_t$ denotes the model’s estimated probability for an input to be classified into class $t$.
Performance was evaluated using segment-based recall (R), precision (P), macro-F-measure (Macro-F), and micro-F-measure (Micro-F)\cite{mesaros2016metrics} using sed\_eval\footnotetext[4]{\url{https://tut-arg.github.io/sed_eval/index.html}}.\yy{ The macro-F-measure and micro-F-measure denote the class-wise and instance-wise average of the F-measure, respectively.}
We set the segment length to 50 ms.

% \subsection{Auxiliary information}
\subsection{Model Improvement}
%\yy{\subsection{Hints to model}}
\subsubsection{Label-frame sparsity}
As mentioned in Section \ref{sec:analysis}, most singing techniques have a short duration (i.e., shorter than 1 s). 
This can cause label imbalance between non-technique frames.
To alleviate the problem, we investigated the effect of \textit{focal loss} \cite{lin2017focal}, which was originally proposed for image object detection.
The focal loss can be represented as follows:
\begin{equation}
    L_{fl}(p_t) = -\alpha (1-p_t)^\gamma \log (p_t)
\end{equation}
where $\alpha \in [0, 1]$ is a weighting factor for balancing the importance of positive and negative examples, and the term $(1-p_t)^\gamma$ is a modulating factor, with $\gamma$ controlling the rate of dominant examples. 
% We set $\alpha=1$ and $\gamma = 2$ for all conditions in the work that used focal loss. 
We set $\alpha=0.2$ and $\gamma = 2$ for all conditions in the work that used focal loss. 
\subsubsection{Pitch information}
\yy{We demonstrated that some of the techniques, such as vibrato and scooping, have a pattern of the shape to certain extent.}
% are pitch modulation or evolution-based techniques. 
Therefore, it is possible that explicitly feeding the pitch helps in the detection.
We further investigated the effect of the auxiliary information.
% as one of the other identification tasks in MIR \cite{hung2018frame,hsieh2020addressing,lordero2021pitch}, which is based on DNN.
Under this condition, we considered two ways of obtaining pitch in our experiment.
One is the ground-truth (GT) pitch annotation mentioned in Subsection \ref{sec:melody}.
However, when used in real-world applications, it is difficult to obtain correct pitch annotation. 
Hence, we also used pitch estimation predicted by CREPE\cite{kim2018crepe} on a separated vocal track. 
As CREPE can compute pitch confidence, we adopted the pitch value where the confidence value was higher than 0.5, whose overall accuracy was 78.0\% evaluated by mir\_eval \cite{raffel2014mir_eval}\footnotetext[5]{The other metrics were following; 85.9\% for voice recall, 29.9\% for voicing false alarms, 94.5\% for raw pitch accuracy.}.
% We note that the performance scores of CREPE on the dataset were 85.9\% for voice recall, 29.9\% for voicing false alarms, 94.5\% for raw pitch accuracy, and 78.0\% for overall accuracy, as evaluated by mir\_eval \cite{raffel2014mir_eval}.
We converted the pitch contour to a mel-band pitchgram that has the same frequency dimensions as the input mel-spectrogram, as in the work of singer identification \cite{hsieh2020addressing}.
Therefore, we stacked it on a mel-spectrogram along the channel axis.

\subsection{Experimental Results}
% The experimental results are listed in Table \ref{tab:exp_results}. 

 \begin{table}[!t]
\centering

\label{tab:exp_results}
\scalebox{0.9}{
\begin{tabular}{|l|l|l|l|l|}
\hline
 & Macro-F & Micro-F  & P & R   \\ \hline
BCE & 37.7\%  & 56.6\% & 40.2\% & 41.7\% \\ \hline
Focal                        & 39.6\%          & 57.7\%          & 40.3\%            & 42.6\%          \\ \hline

BCE-GT & 39.1\%& 57.1\% & 41.4\% & 42.8\%   \\
BCE-CREPE & 39.1\% & 57.1\% &  40.2\% & 41.7\%    \\ \hline

Focal-GT          & \textbf{40.4\%} & \textbf{58.3\%} & \textbf{43.1\%}   & 42.2\%          \\
Focal-CREPE       & 40.3\%          & 57.9\%          & 42.7\%            & \textbf{43.3\%} \\ \hline
\end{tabular}
}
\caption{The results of singing technique detection.}
\end{table}

\subsubsection{Baseline}
\begin{figure}
     \centering
     \includegraphics[width=\columnwidth]{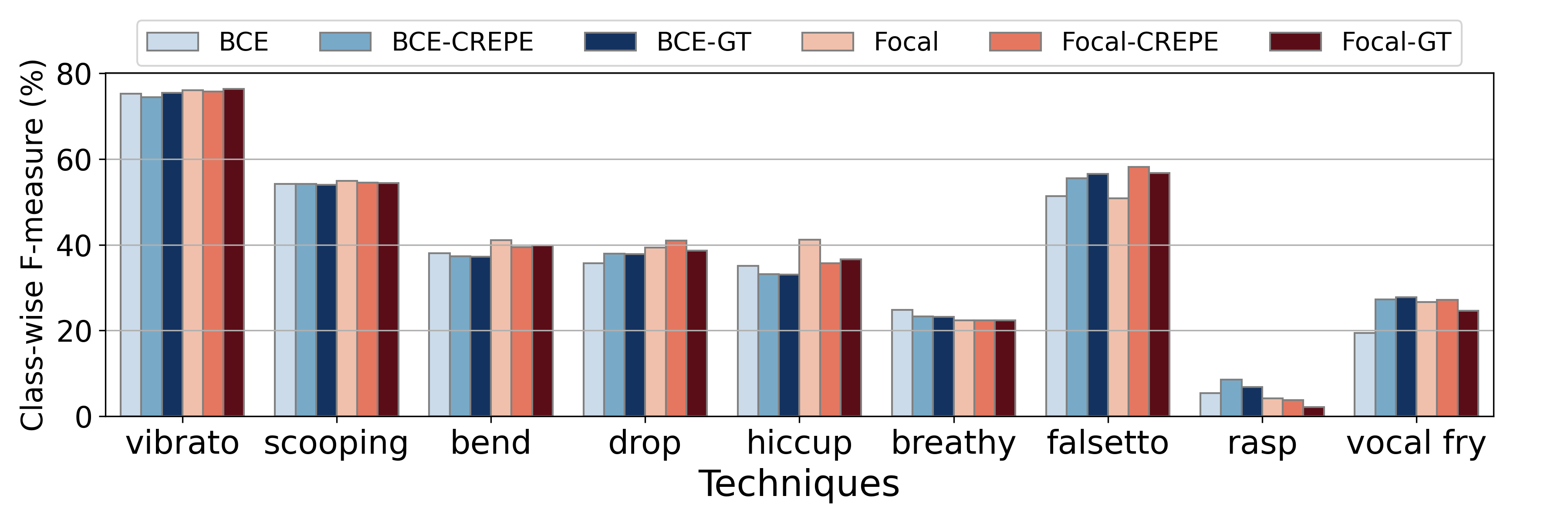}
     \caption{Class-wise F-measure}
     \label{fig:class_wise_result}
 \end{figure}

The first row of Table \ref{tab:exp_results} shows the results of the CRNN model compiled by BCE: 37.7\% for Macro-F and 56.6\% for Micro-F. The fact that Macro-F is much lower than Micro-F indicates that the model failed to identify minority rather than majority classes. Actually, the class-wise F-measure, as shown in Figure \ref{fig:class_wise_result}, indicates that detection of the minority techniques (e.g., `rasp' and `vocal fry') is more difficult than the majority techniques (e.g., `vibrato' and `scooping').

% \subsubsection{Effect of feeding auxiliary information\yy{hints}}
\yy{\subsubsection{Effect of focal loss and pitch information}}
First, we show the results for the effect of \textit{focal loss} in the middle part of Table \ref{tab:exp_results}. 
The detection performance improved by 1.9\% in Macro-F. 
We further studied the class-wise F-measure, as shown in Figure \ref{fig:class_wise_result}, and confirmed that the performance of short techniques, such as  `bend' (38.0\% $\rightarrow$ 41.1\%), `drop' (35.7\% $\rightarrow$ 39.3\%), and `hiccup' (35.1\% $\rightarrow$ 41.2\%) are especially improved.
% We showed that the duration of these techniques is relatively short and that focal loss can adapt to the sparsity of the label.
This indicates that \textit{focal loss} can adapt to the sparsity of the label.

Next, we showed the results of the effect of the auxiliary pitch information in the middle part of Table \ref{tab:exp_results}.
A remarkable finding is that the F-measure of `falsetto' was particularly improved (51.3\% $\rightarrow$ 56.5\%). We also confirmed that the recall value of `falsetto' is raised more (64.7\% $\rightarrow$ 74.1\%) rather than the precision value (45.1\% $\rightarrow$ 48.3\%).
This indicates that pitch information hint at the relationship between the sung pitch value and the occurrence of falsetto (i.e., falsetto can appear only at a higher pitch).
We also confirmed that the input of CREPE pitch shows the similar tendency.

We further investigated the effects of combining these two types of auxiliary information. As shown in the bottom part of Table \ref{tab:exp_results}, the combination of GT pitch and focal loss is the best condition in terms of both macro- and micro-F values. 
We can confirm that both effects of each auxiliary information occur under this condition (red and dark-red bars in Figure \ref{fig:class_wise_result}). 
This indicates that the condition is robust to techniques that have a short duration or some relationships with pitch. 
 
\yy{\subsubsection{Challenges}
The remaining challenges in the task are detected unnatural region lengths and singers' identity.
We confirmed that one of the common mis-detection cases is from the detection of too short or frequently switching regions.
It might need to consider post-processing, such as temporal filtering or probabilistic model (e.g, hidden Marcov model, hidden semi-Marcov model, etc.).
As for singer identity, it can affect the performance of the timbral techniques.
We labeled timbral techniques when the sung voice transformed from the ordinary voice of the singer, and it caused the confusion.
For example, the `rasp' from a singer who has clean voice and ordinary voice from a singer who has raspy voice can be confusing. 
Therefore, there is still an issue of disentangling singers' identity and singing techniques.}

\section{Conclusion and future work}

This work presented an analysis and identification of the singing technique in J-POP music as a case study. In addition, we built a new dataset consisting of 168 J-POP songs with annotation of pitch contour and singing techniques. 
The contributions of this study are summarized as follows: 
1) we constructed a dataset with frame-level annotations of singing techniques on 168 famous J-POP vocal songs, 
2) conducted a descriptive-statistical analysis of the dataset,
3) proposed a CRNN-based singing technique detection model as a baseline, and confirmed that both auxiliary pitch information and focal loss help the detection performance. 

% For future work, we plan to annotate the dataset with notes and conduct a joint analysis between note timing and height, which will accelerate the clarification of the singing style.
\yy{As we showed the appearance of singing techniques relates to some other factors (e.g., difference of singing technique distribution by singer, pitch information helps the detection of falsetto, etc.), we are planning further investigations of relationships or joint identification between other musical factors (e.g. musical note, lyrics, beat and tempo, etc.) or higher level information (e.g. singer, songs' emotion, mood, etc.) with making more types of annotation.
Although this work focuses only on J-POP, the knowledge can be applicable to other types of singing. 
Similar study on other style of vocal such as K-POP, western pops or non-popular vocal and comparison between them are other interesting topics for future work.}

\yy{\section{Acknowledgements}
This work was supported by JST SPRING, Grant Number JPMJSP2124.
Dr. Maiko Murai (Associate Prof. at the University of Tsukuba) gave advices on copyright law.
Yoshihide Ishikawa, Sayori Takayama and Karolina Shi helped the preparation of the metadata.
% The community of IPSJ SIGMUS and ASJ in Japan gave us many helpful feedbacks on ongoing report.
We are grateful to above all.
}
\bibliography{ISMIRtemplate}

\end{document}